\documentclass[letterpaper]{article}
\usepackage{aaai24} %
\usepackage{times} %
\usepackage{helvet} %
\usepackage{courier} %
\usepackage[hyphens]{url} %
\usepackage{graphicx} %
\usepackage{graphicx}  %
\usepackage{natbib}  %
\usepackage{caption}  %
\usepackage{subcaption}
\usepackage{url}
\usepackage{booktabs}
\usepackage[most]{tcolorbox}
\usepackage{amssymb}
\usepackage{colortbl}
\usepackage{xtab,afterpage}
\usepackage{svg}
\usepackage[most]{tcolorbox}
\usepackage{float}

\title{Gender, Race, and Intersectional Bias in Resume Screening via Language Model Retrieval}
\author{
Kyra Wilson, Aylin Caliskan\\
}
\affiliations {
University of Washington Information School\\
Seattle, Washington, USA\\
\{kywi, aylin\}@uw.edu\\
}
\date{July 2024}

\begin{document}
\frenchspacing

\maketitle

\begin{figure}
    \begin{tcolorbox}[colback=red!5!white,colframe=red!75!black]
    August 29, 2026: After attempting to replicate our study, Frank Pallas (frank.pallas@plus.ac.at, Salzburg University) and Martin Wiesinger (martin.wiesinger@plus.ac.at, Salzburg University) informed us that our published results are inaccurate due to a bug in our code. Specifically, the results of \textbf{gender-only} comparisons should be inverted, such that measurements referring to the preference of men should actually refer to preference of women, and vice versa. The results for \textbf{race-only} and \textbf{intersectional} comparisons are unaffected by this error and were replicated by the Salzburg University team. We are grateful to them for bringing this error to our attention.
    \end{tcolorbox}
\end{figure}

\begin{abstract}
Artificial intelligence (AI) hiring tools have revolutionized resume screening, and large language models (LLMs) have the potential to do the same. However, given the biases which are embedded within LLMs, it is unclear whether they can be used in this scenario without disadvantaging groups based on their protected attributes. In this work, we investigate the possibilities of using LLMs in a resume screening setting via a document retrieval framework that simulates job candidate selection. Using that framework, we then perform a resume audit study to determine whether a selection of Massive Text Embedding (MTE) models are biased in resume screening scenarios. We simulate this for nine occupations, using a collection of over 500 publicly available resumes and 500 job descriptions. We find that the MTEs are biased, significantly favoring White-associated names in 85.1\% of cases and female-associated names in only 11.1\% of cases, with a minority of cases showing no statistically significant differences. Further analyses show that Black males are disadvantaged in up to 100\% of cases, replicating real-world patterns of bias in employment settings, and validate three hypotheses of intersectionality. We also find an impact of document length as well as the corpus frequency of names in the selection of resumes. These findings have  implications for widely used AI tools that are automating employment, fairness, and tech policy.
\end{abstract}

\section{Introduction}
One of the widespread practical applications of artificial intelligence (AI) tools has been their use in hiring processes. It is estimated that 99\% of Fortune 500 companies are already using some sort of AI assistance when making hiring decisions \citep{Schellmann2024-as}, due to their potential to increase recruitment quality and efficiency \citep{chen2023ethics}. This could potentially alleviate discrimination based on people's unconscious biases or stereotypes, such as associating Black male job seekers with criminals \citep{pager2003mark} or female job seekers with lower productivity due to motherhood \citep{gonzalez2019role}. However, many AI hiring tools do still exhibit biased outcomes, such as a resume screening tool developed at Amazon which had to be scrapped when it was revealed that it unfairly discriminated against women \citep{dastin2018}.

Large language models (LLMs), which are trained on a general corpus of language data rather than data specific to hiring tasks, also have the potential to be used in these scenarios. In fact, resume screening tasks have already been observed in interactions between users and ChatGPT \citep{ouyang-etal-2023-shifted}. Additionally, the accessibility of these models (both in terms of cost and user interfaces) lend them well to adoption by companies which have either not yet incorporated AI assistance into their hiring pipeline due to cost or technological complexity or to replace models that are currently in use and whose biases may be better understood.  

With the potential for increased use of LLMs in hiring, it is essential to document the extent to which they exhibit biases against particular social groups. In the US, it is illegal to make hiring decisions on the basis of race, color, religion, sex (including gender identity, sexual orientation, and pregnancy), national origin, age (40 or older), disability or genetic information. A number of these biases have been documented in LLMs already, including gender, race, religion, and disability biases, as well as their intersections \citep{kotek, nvenkit, kirk2021bias, guo2021detecting}. Therefore, it is crucial to investigate whether these models exhibit discriminatory biases related to protected attributes such as race and gender or their intersections in order to evaluate how they can  be used for resume screening tasks. 

Additionally, it is also essential to investigate low-level textual features such as term frequencies and document lengths which, although not direct signals of social groups, play a significant role in the performance and outputs of language models \citep{jones2022capturing, anil2022exploring, wolfe2021low}. These features can vary widely across documents such as resumes, and evaluating their relationship to social group outcomes in automated resume screening is essential in order to accurately represent real-world usage.

In this study, we formulate resume screening as a practical zero-shot document retrieval task. Using this approach, we seek to address the following questions:
\begin{itemize}
  \item \textbf{RQ1:} Are identical resumes with different race (Black vs. White) or gender (male vs. female) signals selected at equal rates when using three state-of-the-art LLMs for resume screening via a practical retrieval task?
  \item \textbf{RQ2:} Are identical resumes with different intersectional group signals selected at equal rates when using three language models for the same resume screening task?
  \item \textbf{RQ3:} How do the features of race and gender signals such as name frequency and resume length impact screening outcomes?
\end{itemize}

We introduce a simulation for LLMs' usage as resume screening tools and analyze outcomes with respect to race and gender. In a hiring decision pipeline, resume screening is generally considered to be the second stage, after sourcing potential candidates, and prior to interviewing and selecting a candidate \citep{bogen2018help}. LLM assistance is particularly useful for this stage because it involves the analysis of many text documents (resumes) to identify those that are most relevant relative to a particular job description. 

We investigate LLM-mediated resume screening using publicly-available English datasets of over 500 resumes and 500 job descriptions that are collected from real-world examples of these documents. They span nine occupations (Chief Executive, Marketing and Sales Manager, Miscellaneous Manager, Human Resources Worker, Accountant and Auditor, Miscellaneous Engineer, Secondary School Teacher, Designer, and Miscellaneous Sales and Related Worker). Resume screening is analyzed within three different Massive Text Embedding (MTE) models, a special class of LLMs which are trained for representational tasks such as document retrieval, classification, and clustering after pre-training on general language corpora. While many studies have characterized the biases of foundation or instruction-tuned LLMs, very few have investigated the biases of MTEs or their use in resume screening, adding further novelty and importance to this study. 

To audit the MTE models for biases in resume screening scenarios, we augment resumes with 120 frequency-controlled names that are associated with White, Black, male, and female identities. Using document retrieval and selecting subsets of resumes which are most similar to job descriptions, we are able analyze whether outcomes differ across identity groups. The code and data produced by this research are available at \url{https://github.com/kyrawilson/Resume-Screening-Bias}. We make the following knowledge contributions:

\begin{enumerate}
    \item In all of the models, we find that resumes that belong to the same occupation category as a given job description have significantly higher cosine similarities than resumes that belong to a different occupation category (0.0437 higher on average), justifying the use of cosine similarity to  determine retrieval in order to analyze an initial stage of resume screening where a pool of the most relevant candidates are identified for further evaluation by a hiring professional.
    \item Using more than three million comparisons between resumes and job descriptions, we find that resumes with White or male names are preferred to those with Black or female names in up to 85.1\% of cases.
    \item Intersectional comparisons reveal resumes that contain Black male names are highly unfavored in resume screening, with other groups being preferred in up to 100\% of cases. Gender differences are driven largely by disparate preferences of Black females over Black males, as White males and White females have much smaller selection rate differences. These findings validate three hypotheses of intersectionality.
    \item Features such as resume length and name frequency significantly impact bias measurements in LLM resume audits, such that increasing the ratio of signals that are proxies to race or gender information in a document by decreasing its length can increase the number of biased outcomes by  22.2\%, and changing frequency matching strategies can alter whether Black names or White names are favored in a majority of cases.
\end{enumerate}

Finally, we discuss how the resume screening patterns found in LLMs echo societal patterns of hiring discrimination and how the particular features of resumes such as names should be considered when using LLMs for resume screening.

\section{Related Work}
There has been limited work addressing and documenting the potential risks of using LLMs for hiring decisions, despite the documentation of biases in real-world resume screening scenarios and in AI models specially trained for hiring tasks. In economic labor market studies, bias in resume screening is usually investigated through resume audit or correspondence studies \citep{baert2018hiring}. In these experiments, artificial resumes which differ only on some protected attribute are created and then sent in response to real job postings. Hiring procedures are determined to be discriminatory if the response rates differ significantly between groups that vary on the key dimension. This paradigm has been used to identify bias related to a number of protected attributes such as race \citep{bertrand2004emily}, queerness \citep{mishel2016discrimination}, religion \citep{wright2013religious}, disability \citep{hipes2016stigma}, and age \citep{neumark2016experimental, neumark2019harder, lahey2008age}.

To date, there are no external gender or racial bias audits of AI-mediated resume screening tools, the majority of which are typically closed-source, propriety, and not accessible for external review \citep{li2024making}. Limited work has addressed this issue by reviewing publicly available statements and model descriptions \citep{raghavan2020, ukhiring}, finding that the majority of vendors do not make explicit statements regarding their models' compliance with anti-discrimination law, and those that do are typically only within a US context. \citet{pymetrics}, using a cooperative audit, found that the system of interest did not exhibit adverse impact, but key assumptions make this result difficult to generalize without additional testing. A final external audit found that closed-source models are often unstable, but they did not investigate any protected characteristics \citep{crystal}.

Only very preliminary work has been done to investigate LLMs used for resume screening. A team of Bloomberg reporters investigated OpenAI's GPT-3.5 and GPT-4 and found that Black women were only ranked as top candidates for software engineering roles in 11\% of tests, and Hispanic women were twice as likely as men to be ranked as top candidates for human resources workers. The only career in which no group was disadvantaged was retail workers \citep{Yin_Alba_Nicoletti_2024}. Another study investigating OpenAI's ChatGPT found that resumes which mentioned disability in the context of an award were only ranked highest in 25\% of cases \citep{glazko2024identifying}. While both studies demonstrate biased outcomes when using LLMs or chatbots as resume screeners, the models investigated were all ``black boxes," meaning the analysis was limited to model outputs only and could not investigate internal representations. Additionally, researchers did not rigorously investigate low-level document features such as term frequency or length in their studies, which are related to model biases \citep{esiobu-etal-2023-robbie}. 

Another limitation of studies investigating chatbots' use as resume screeners is that they are less transparent and rely on the model's generative capabilities for resume screening. Language generation is relatively computationally expensive as new tokens are produced iteratively, and outputs are also highly sensitive to features of the prompt which are unrelated to the task itself \citep{sclar2023quantifying}. An alternative is to use embeddings of documents directly, which are less computationally intensive to process although still potentially sensitive to spurious prompt features.

\section{Data}
\begin{figure}
\centering
\includegraphics[width=0.4\textwidth]{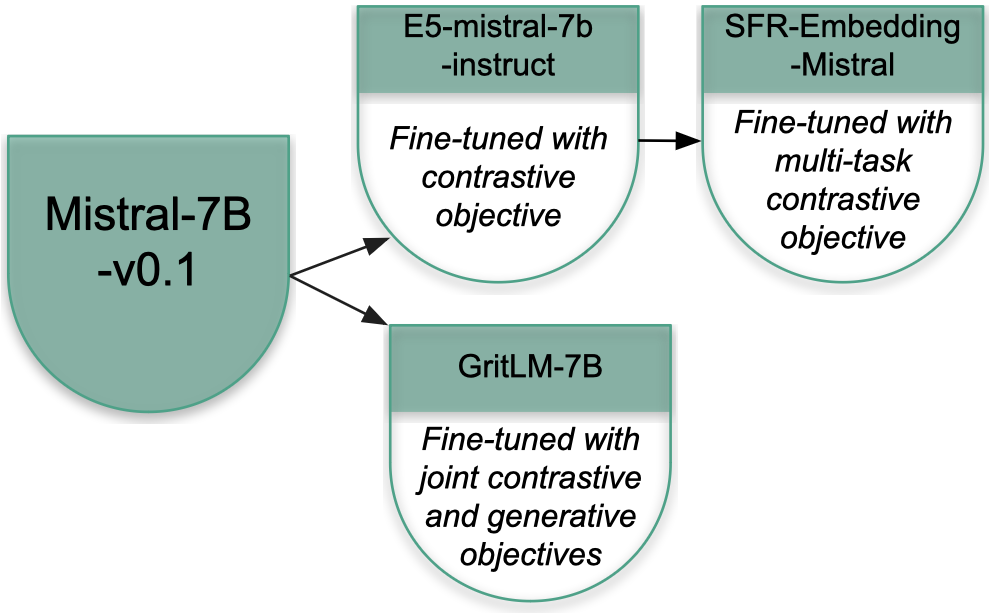}
\caption{Relation of the three models investigated to a pre-trained LLM. Arrows between models indicate additional fine-tuning steps.}
\label{mistral-flow}
\end{figure}

The present study examines race and gender bias in resume screening using a corpus of resumes and job descriptions as well as three MTEs (illustrated in Figure \ref{mistral-flow}) with a Mistral-7B-like architecture \citep{jiang2023mistral}. Mistral-7B is based on the transformer architecture \citep{vaswani2017attention}, but it innovates using grouped-query attention and sliding-window attention to decrease computational costs while also being able to process long sequences more effectively, a key innovation for the resume screening setting where long documents are common. 

\subsection{Massive Text Embedding Models}
We select three high-performing MTEs to illustrate the potential range in outcomes from a set of architecturally similar models. The chosen models were fine-tuned for representational tasks (including document retrieval, classification, and clustering) on either Mistral-7B-v0.1, a foundation model whose learning objective is solely next-token prediction, or on models which were themselves fine-tuned on Mistral-7B-v0.1. In order to achieve high performance on these tasks, LLMs must undergo an additional round of training (\textit{fine-tuning}) in which a specialized loss function is used to update model parameters. Typically, the loss function used for representational tasks is a contrastive loss (CL), in which embeddings of examples that are positive examples of a query or label are brought closer together spatially, while those that are negative examples are pushed further apart \citep{jones2022capturing}. 

E5-mistral-7b-instruct (\textit{e5}) \citep{wang2023improving} is trained on Mistral-7B-v0.1 and fine-tuned using using a CL objective with natural and synthetically generated training data for retrieval tasks. GritLM-7B (\textit{GritLM}) \citep{muennighoff2024generative} is also fine-tuned on top of Mistral-7B-v0.1; however, it has no synthetic data in its training set and uses a joint objective with both CL and next-token prediction in order to train for both representational and generative functions. Finally, SFR-Embedding-Mistral (\textit{SFR}) \citep{SFRAIResearch2024} is fine-tuned with CL on top of e5, which itself is already a fine-tuned MTE. Its training data includes both retrieval tasks as well as classification and clustering tasks. 

All three MTEs achieve state-of-the-art performance according to the Massive Text Embedding Benchmark (MTEB), which aims to quantify the performance of MTEs on a large set of representational tasks and datasets \citep{muennighoff2022mteb}. On document retrieval tasks, they are the highest performing open-source MTEs with at least one billion parameters\footnote{As of April 2024.}.

\begin{table*}
\small
\centering

\begin{tabular}{|c|c|c|c|c|c|c|}
\hline
\textbf{Broad Occupation} & \textbf{\%Women} & \textbf{\%White} & \textbf{\%Black} & \textbf{\begin{tabular}[c]{@{}c@{}}Total Workers \\ (Thousands)\end{tabular}} & \textbf{\begin{tabular}[c]{@{}c@{}}\# of \\ Resumes\end{tabular}} & \textbf{\begin{tabular}[c]{@{}c@{}}\# of \\ Descriptions\end{tabular}} \\ \hline
Chief Executives          & 30.6            & 85.8            & 5.2             & 1,780                                                                         & 49                                                               & 33                                                                    \\ \hline
Marketing and             &                 &                 &                 &                                                                               &                                                                  &                                                                       \\
Sales Managers            & 47.7            & 86.9            & 5.0             & 1,136                                                                         & 51                                                               & 81                                                                    \\ \hline
Miscellaneous             &                 &                 &                 &                                                                               &                                                                  &                                                                       \\
Managers                  & 37.5            & 80.7            & 9.2             & 5,666                                                                         & 112                                                              & 92                                                                    \\ \hline
Human Resources           &                 &                 &                 &                                                                               &                                                                  &                                                                       \\
Workers                   & 76.5            & 74.9            & 14.7            & 980                                                                           & 22                                                               & 27                                                                    \\ \hline
Accountants and           &                 &                 &                 &                                                                               &                                                                  &                                                                       \\
Auditors                  & 57.0            & 73.4            & 11.9            & 1,624                                                                         & 133                                                              & 99                                                                    \\ \hline
Miscellaneous             &                 &                 &                 &                                                                               &                                                                  &                                                                       \\
Engineers                 & 15.4            & 72.4            & 5.9             & 669                                                                           & 28                                                               & 112                                                                   \\ \hline
Secondary School          &                 &                 &                 &                                                                               &                                                                  &                                                                       \\
Teachers                  & 56.9            & 87.8            & 6.1             & 944                                                                           & 20                                                               & 52                                                                    \\ \hline
Designers                 & 52.0            & 80.5            & 6.6             & 862                                                                           & 87                                                               & 55                                                                    \\ \hline
Miscellaneous Sales       &                 &                 &                 &                                                                               &                                                                  &                                                                       \\
and Related Workers       & 56.1            & 81.0            & 12.0            & 339                                                                           & 52                                                               & 20                                                                    \\ \hline
\end{tabular} 
\caption{Resume screening was investigated for nine occupations. These are presented with corresponding US population statistics for percentage of women workers, White workers, Black workers, and total number of workers as well as the number of documents corresponding to each occupation category from the resume and job description datasets after filtering.}
\label{resume-table}
\end{table*}

\subsection{Job Description and Resume Stimuli}
In order to simulate resume screening, a selection of job descriptions as well as candidate resumes are necessary. While examples of job descriptions are widely available online, they are frequently removed or altered as positions fill or requirements change. This makes reproducing research using these descriptions difficult. Additionally, resumes contain a wealth of sensitive and private information, so can be difficult to both collect and disseminate. To ensure reproducibility, we select two job description and resumes datasets which are publicly available online\footnote{Resumes were collected from \url{https://www.kaggle.com/datasets/snehaanbhawal/resume-dataset}. Job descriptions were collected from \url{https://www.kaggle.com/datasets/marcocavaco/scraped-job-descriptions}.}. Job descriptions are a free-text solicitation for an open position; resumes have an occupation title followed by free-text description of qualifications.

\begin{figure*}[ht]
\centering
\includegraphics[width=0.7\textwidth]{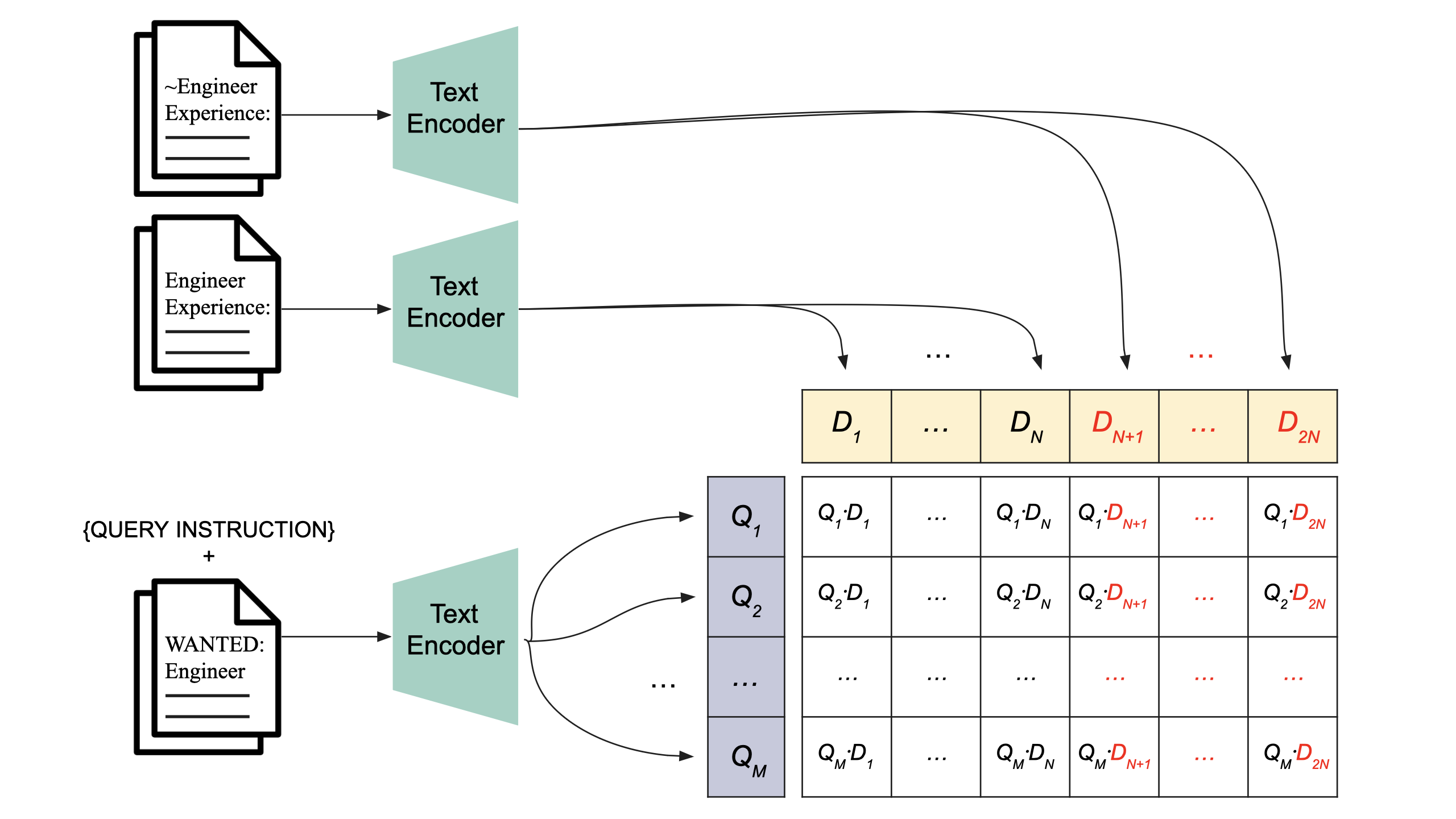}
\caption{Illustration of the resume screening as document retrieval framework. Task instructions are appended to job descriptions and treated as queries, while resumes are treated as documents. The cosine similarity between queries and documents estimates the relevance of a resume to a particular job description. }
\label{doc-retrieval}
\end{figure*}

Because these datasets were not annotated with the same occupation classification schemes, it was necessary to standardize the job titles associated with each document in the dataset. The US government developed the Standard Occupational Classification (SOC) system in order to classify and collect data about particular occupations. Using this scheme, six-digit codes were assigned for each job description and resume using the NIOSH Industry and Occupation Computerized Coding System (NIOCCS)\footnote{https://csams.cdc.gov/nioccs/} from the US government. This tool was developed to programatically assign SOC codes to free text job titles and descriptions\footnote{\citet{schmitz2016industry} investigated the NIOCCS with respect to manual coding and found that the tool had ``fair" to ``good" accuracy when assigning codes with up to four digits. Although this is lower than what is reported by NIOSH, we find that this tool still has superior performance on our sample of resumes and job descriptions in comparison to other SOC coding tools.}. Code assignments with less than 60\% confidence were eliminated from that dataset. This threshold was chosen to maximize code confidence while also maintaining a large dataset with a range of occupations represented. 

Using the broad occupation (first five digits of the SOC code) identified for each resume and job description, we filtered the dataset to remove duplicates and keep only documents belonging to occupation categories which had at least 20 resumes and 20 job descriptions, ensuring a large pool of documents for accurately simulating resume screening and achieving necessary statistical power. After this, the dataset contained 554 resumes and 571 job descriptions across nine occupations (Chief Executive, Marketing and Sales Manager, Miscellaneous Manager, Human Resources Worker, Accountant and Auditor, Miscellaneous Engineer, Secondary School Teacher, Designer, and Miscellaneous Sales and Related Worker). A complete listing of the categories and associated employment statistics from the Bureau of Labor Statistics (BLS) can be seen in Table \ref{resume-table}. 

\section{Approach}
The present study describes a generalizable and scalable approach to resume screening based on document retrieval. This involves using MTEs to create embeddings for job descriptions and resumes, and then using a simple cosine similarity comparison to capture which resumes are most similar to a given job description. A summary of this approach is given in Figure \ref{doc-retrieval}. 
A chi-square test is then used to determine whether the most similar resumes are distributed equally amongst relevant groups, or whether certain groups are favored over others, indicating bias. Chi-square tests require a minimum of five observed values for valid population estimates \citep{franke2012chi}; our dataset far exceeds that, with at least 160 resume documents for every bias test, demonstrating the legitimacy of the results at scale. 

\subsection{Task-Augmented Job Descriptions}

In document retrieval with MTEs, query texts are encoded with an additional instruction describing the particular setting.\footnote{The text formatting used to encode instructions and query texts varies between MTEs. For each model, we followed the recommended structure as described in that model's documentation for these experiments.} We created a set of 10 instructions, shown in Table \ref{instructions}, to be encoded with job descriptions according to templates specific to each model. We used ChatGPT in this procedure to develop alternatives for the phrases ``job description" and ``resume." 

\begin{table}[t]
\small
\centering
\begin{tabular}{|l|c|c|c|}
\cline{1-4}
\multicolumn{1}{| c}{  \textbf{Name}} & \multicolumn{1}{c}{\textbf{\begin{tabular}[c]{@{}c@{}}Lg. Freq.\end{tabular}}} & \multicolumn{1}{c}{\textbf{\begin{tabular}[c]{@{}c@{}}Distinct\end{tabular}}} & \textbf{Group} \\ \cline{1-4} \cline{1-4}
Kenya                             & 16.87                                                                                   & 1.80                                                                                     & BF             \\ 
Ebony                             & 15.18                                                                                   & 1.29                                                                                     & BF             \\ 
Latrice                           & 10.86                                                                                   & 0.71                                                                                     & BF             \\ 
Latisha                           & 10.77                                                                                   & 1.57                                                                                     & BF             \\ \cline{1-4}
Jackson                           & 17.71                                                                                   & 0.67                                                                                     & BM             \\
Abdul                             & 16.18                                                                                   & 0.82                                                                                     & BM             \\
Demetrius                         & 13.14                                                                                   & 0.89                                                                                     & BM             \\
Dewayne                           & 12.01                                                                                   & 0.97                                                                                     & BM             \\ \cline{1-4}
May                               & 19.45                                                                                   & 0.66                                                                                     & WF   \\ 
Hope                              & 17.86                                                                                   & 1.07                                                                                     & WF  \\ 
Stacy                             & 14.85                                                                                   & 1.37                                                                                     & WF   \\ 
Kristine                          & 13.73                                                                                   & 1.30                                                                                     & WF   \\ \cline{1-4}
John                              & 19.22                                                                                   & 0.89                                                                                     & WM   \\
Joe                               & 17.90                                                                                   & 1.14                                                                                     & WM   \\
Stevie                            & 14.88                                                                                   & 0.66                                                                                     & WM  \\
Huey                              & 13.73                                                                                   & 0.69                                                                                     & WM \\ \cline{1-4}

Virginia                          & 17.84                                                                                   & 1.05                                                                                     & =WF           \\ 
Katie                             & 16.18                                                                                   & 1.66                                                                                     & =WF           \\ 
Aileen                            & 13.12                                                                                   & 0.87                                                                                     & =WF           \\ 
Rebeca                            & 11.97                                                                                   & 1.30                                                                                     & =WF           \\ \cline{1-4}
Daniel                            & 17.71                                                                                   & 1.06                                                                                     & =WM           \\
Spencer                           & 16.18                                                                                   & 0.91                                                                                     & =WM           \\
Bennie                            & 12.95                                                                                   & 0.70                                                                                     & =WM           \\
Wilbert                           & 12.317                                                                                  & 0.87                                                                                     & =WM           \\ \cline{1-4}

\end{tabular} 
\caption{The two most and least frequent first names used for each intersectional group along with their corpus frequencies and racial distinctiveness scores. $=$ indicates White names which are matched exactly to the frequencies of the Black names rather than using names whose frequencies are proportional to US population differences.}
\label{names-main}
\end{table}

\subsection{Name-Augmented Resumes}
To measure bias in resume screening, resumes were augmented with a name, comprised of a variable first name and constant last name, by prepending the complete name to the beginning of the document. \textit{Williams} was selected as a last name because it is both frequent (third most common name in the US) and approximately equally likely to be used either by a Black or White person (47.68\% vs. 45.75\%) \citep{namecensusWhat5000}. The last name was kept constant across all resumes in order to maximize experimental control and document realism while also minimizing required computation.

We use the name database introduced in \citet{elder2023signaling} to select names associated with one of four groups: Black males, Black females, White males, or White females. Of these, the Black male group had the fewest potential names, and the top 20 most distinctive\footnote{Distinctiveness was measured via the difference in ratings from one to five between the most likely racial group versus the second most likely racial group, with all names having a score of at least 0.66.} names (33\% of all Black male names in the database) were chosen for use in resume augmentation.  

An equal number of names corresponding to other groups were then selected in order to closely match or be proportional to the corpus frequencies of the Black male names. Corpus frequencies were determined using infini-gram \citep{Liu2024InfiniGram}, a tool that facilitates n-gram searches for arbitrarily large corpora, and the DOLMA corpus \citep{dolma}\footnote{Although Mistral model weights are available publicly, the training dataset is proprietary. The DOLMA frequencies are meant only to approximate the frequencies of names in the corpus used to train Mistral. DOLMA contains 3.1 trillion tokens and is currently the largest available to search using infini-gram.}. 

The first set of names was selected in order to be proportional to the relative population differences between Black and White people in the US, replicating the distribution of names that would likely be seen in real-world resume screening. According to 2023 US Census estimates, those who identify as White alone comprise 75.5\% of the US population, while those who identify as Black alone comprise 13.6\%. Accordingly, we selected White male and female names which were approximately 5.5 times more frequent in the corpus than corresponding Black male names, and Black female names which were approximately equally frequent to Black male names. We created an additional set of 20 White male and 20 White female names which were also approximately equally frequent to Black male names for supplemental investigation of frequency effects. A sample of first names from both the proportional frequency-matched and exact frequency-matched sets can be seen in Table \ref{names-main} with a full list available in the Appendix.\footnote{A long-form version of this paper with Appendix is available at \url{https://arxiv.org/abs/2407.20371}.}

\subsection{Resume Screening}

Zero-shot dense retrieval, which uses contextualized embeddings to compare documents rather than exact term matches, provides a natural analog for resume screening. In the initial stages of retrieval, relevance scores computed from text embeddings are used to select a set of documents from a large corpus that best match a user query. Cosine similarity is commonly used as a relevance metric \citep{zeroshot}. Similarly for resume screening, resumes $r$ which are similar to a job description $d$ can be identified via their respective embeddings $v_{r}$ and $v_{d}$ using the equation in (\ref{cos-sim}). Furthermore, using a retrieval approach for resume screening allows for the direct analysis of textual embeddings to determine whether the representations are potentially biased in a way that could influence model outputs. If the resumes which are most similar to a particular job description consistently belong to a certain group, this is evidence that the representations are biased in favor of that group. 

\small
\begin{equation}
    sim(r, d) = \langle v_{r}, v_{d} \rangle
    \label{cos-sim}
\end{equation}
\normalsize

Embeddings for document retrieval were generated using the MTEs in Figure~\ref{mistral-flow}. Texts were truncated due to computational limitations, and 1,300 tokens was chosen as a length which captured the majority of resume content while still being computationally feasible. A summary of unaltered document lengths is available in the appendix. Document embeddings were extracted from the last hidden state of a model and normalized before computing their cosine similarity. Cosine similarity scores were averaged over task instructions, so the updated similarity computation is as in Equation (\ref{cos-sim-avg}), where $t$ corresponds to the index of the task instructions in Table \ref{instructions} used to form job description embeddings. For completeness, results for individual task instructions are also provided in the Appendix.

\small
\begin{equation}
    sim(r, d) = \frac{1}{10}\sum_{t=1}^{10}{\langle v_{r}, v_{d_{t}} \rangle}
    \label{cos-sim-avg}
\end{equation}
\normalsize

To simulate candidate selection, we select a percentage of the most similar of resumes for each job description for further analysis. A chi-square test is used to determine whether the selected resumes are distributed uniformly amongst relevant groups or whether particular groups are represented at significantly higher rates than others, indicating bias in resume screening outcomes. Results for resume screening outcomes are presented primarily in terms of difference in selection rates; intermediate cosine similarity results are provided in the Appendix. 

\begin{figure}
\includesvg[width=0.465\textwidth]{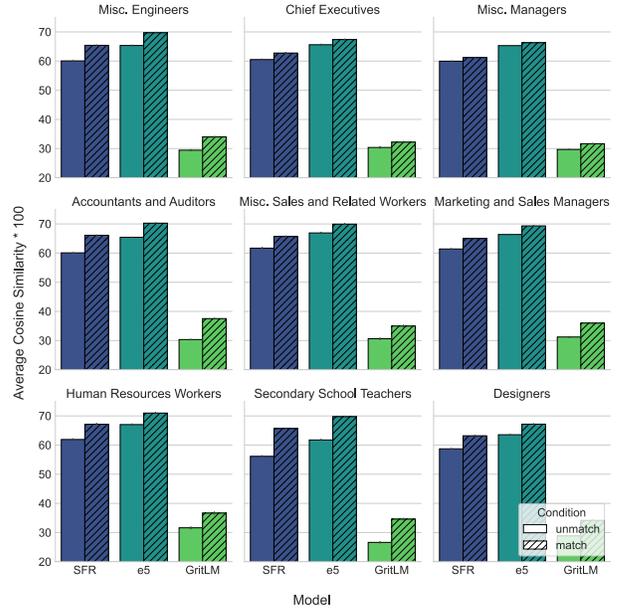}
\caption{For each occupation and model, cosine similarities are significantly higher (p$<$0.001) for resumes which belong to the same occupation as the job description (\textit{match}) than those that belong to different occupations (\textit{unmatch}), indicating the success of the document retrieval for resume screening framework.}
\label{validation}
\end{figure}

\section{Experiments}

Detailed information is provided for four experiments. Experiment 1 evaluates the document retrieval for resume screening framework by comparing the similarity of resumes which belong to the same occupation category as a given job description to those that belong to different categories. Experiment 2 and 3 investigate bias in resume screening, first using gender and race categories separately, then investigating intersectional identities. Finally, Experiment 4 examines the effect of low-level features such as name frequency and resume length on bias measurements. In all experiments, the initial pools of resumes are balanced with respect to identity groups, and the expected outcome is that all groups should be represented uniformly if the MTEs are unbiased. Any significant deviations from this represent biased outcomes against particular identity groups in resume screening. 

\subsection{Evaluating Retrieval for Resume Screening}
Cosine similarity of embeddings for resumes without names and job description embeddings was calculated for both resumes whose occupation category corresponded to the job description (\textit{matched}) and those which did not (\textit{unmatched}), simulating the initial stage of screening resumes for relevance. The cosine similarity scores of these two groups were compared to verify that the document retrieval approach is suitable for resume screening with LLMs.

\subsection{Evaluating Race and Gender Bias}
Gender and race groups were formed by combining names with population proportional frequencies from the four intersectional groups into four groups corresponding to only one race or gender identity (Black, White, male, or female). Each group was comprised of 40 names. Embeddings for job descriptions and name-augmented resumes were created using the three MTE models and cosine similarities were computed.

For each model and occupation, we performed a bias test by selecting the top 10\% of most similar resumes for every job description and determining whether race or gender groups were represented at significantly higher rates. At this threshold, a minimum of 160 resumes were selected for each job description, and a total of 27 bias tests were conducted for both gender and race.

\subsection{Evaluating Intersectional Bias}
Using the 20 names with population proportional frequencies from each intersectional group (Black female, Black male, White female, White male), we repeated the embedding procedures, selection of top 10\% of resumes, and 27 chi-square bias tests from Experiment 2 for each pair of intersectional identities, excluding those in which no race or gender dimension was shared.

\subsection{Evaluating Effects of Length and Frequency}

An additional set of name-augmented resumes was created in which the document contained a name and occupation title only (\textit{title-only}) in contrast to the original name-augmented resumes which contained a name, occupation title, and additional content (\textit{full-length}). Cosine similarities and bias tests were repeated as described above, and the results of using \textit{title-only} versus \textit{full-length} resumes were compared for non-intersectional race and gender categories. 

Finally, the effect of using raw frequency matched versus population proportional frequency matched names was investigated by repeating the cosine similarity and bias tests as described above for resumes augmented with both sets of names, and comparing the results for non-intersectional race and gender categories.

\section{Results}

First, evidence indicates successful resume screening via retrieval. Additional evidence indicates underlying biases favoring resumes with White or male names when race and gender are analyzed independently. Intersectional analyses indicate this bias is strongest against resumes with Black female or male names. Finally, attributes such as name frequency and resume length also effect the similarity of resumes to job descriptions. 

\subsection{Verification of Retrieval}
Resumes with no names belonging to the same occupation category as the job description have significantly higher cosine similarities (p$<$0.001) than resumes with no names belonging to different occupation categories for every occupation, as seen in Figure \ref{validation}, verifying the use of document retrieval for resume screening. 

\begin{figure}[t]
\includesvg[width=0.465\textwidth]{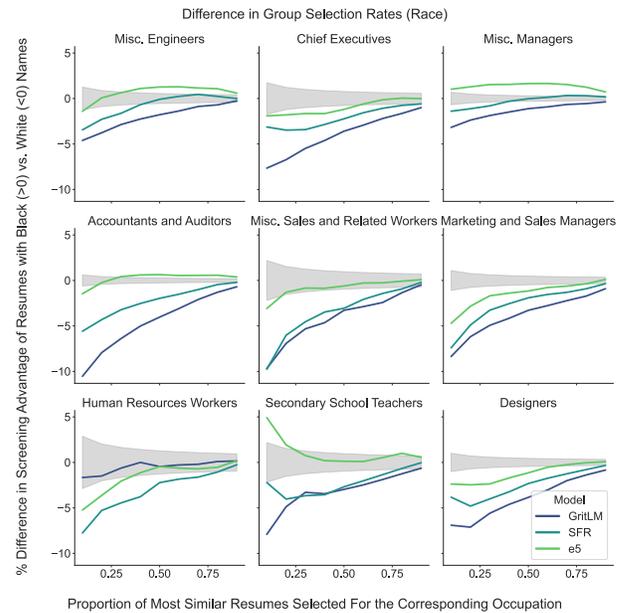}
\caption{Resumes with White names are significantly preferred (p$<$0.05) in 85.1\% of tests; those with Black names are preferred in 8.6\% of tests. Gray regions indicate disparities which are not significantly different from zero (6.3\% of tests).}
\label{race-full}\hfill
\end{figure}

\begin{figure}[t]
\includesvg[width=0.465\textwidth]{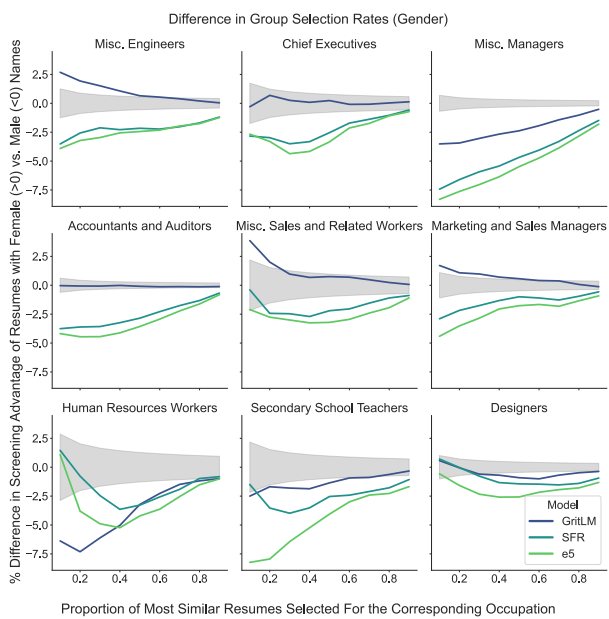}
\caption{Resumes with male names are significantly preferred (p$<$0.05) in 51.9\% of tests; those with female names are preferred in 11.1\% of tests. Gray regions indicate disparities which are not significantly different from zero (37\% of tests).}
\label{gender-full}
\end{figure}

\subsection{White and Male Names are Preferred}

The majority of experiments reveal a preference for White names over Black names. In 85.1\% of 27 tests for racial bias in resume screening, White names were preferred, and only in 8.6\% of tests were Black names preferred, as seen in Figure \ref{race-full}. Notably, the only cases in which Black-associated names were preferred at any threshold were results from the e5 model, suggesting that this MTE may be less racially biased than others for particular occupations.

While male names were also favored compared to female names in the majority of experiments, the disparities were less than those demonstrated using Black versus White names. In 51.9\% of 27 tests for gender bias in resume screening, male names were preferred to female names, and only in 11.1\% of tests were female names preferred, as seen in Figure \ref{gender-full}. Again, all cases in which female names were preferred came from a single model, GritLM.  

\subsection{Nuances of Intersectional Identities}
\begin{figure}[t]
\includesvg[width=0.465\textwidth]{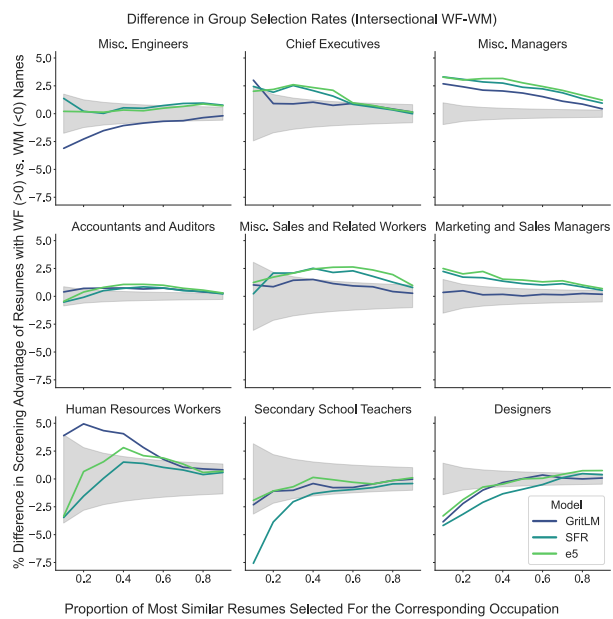}
\caption{Resumes with White male names are preferred in 18.5\% of tests; those with White female names are preferred in 25.9\%. Gray regions indicate disparities which are not significantly different from zero (55.6\% of tests).}
\label{WF-WM}
\end{figure}

Intersectional comparisons reveal that the smallest disparities exist between White names of different genders. Comparisons between Black names of different genders or Black and White names of the same gender exhibit larger disparities. Comparisons between resumes with White male and White female names reveal significant differences (p$<$0.05) between the two groups in only 44.4\% of 27 tests, as shown in Figure \ref{WF-WM}. White males are only preferred in 18.5\% of tests and White females are preferred in 25.9\%. No model exhibits consistent behavior in preferring one group where other models do not. 

\begin{figure}[t]
\includesvg[width=0.465\textwidth]{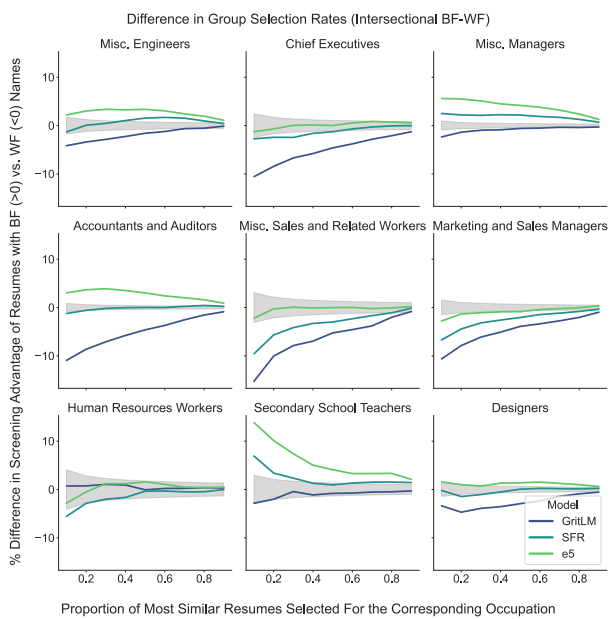}
\caption{Resumes with White female names are preferred in 48.1\% of tests; those with Black female names are preferred in 25.9\%. Gray regions indicate disparities which are not significantly different from zero (26\% of tests).}
\label{BF-WF}
\end{figure}

\begin{figure}[t]
\includesvg[width=0.465\textwidth]{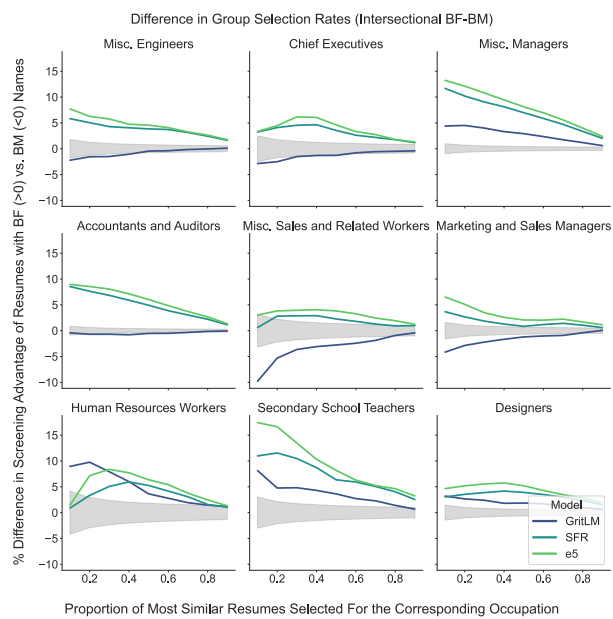}
\caption{Resumes with Black female names are preferred in 66.7\% of tests; those with Black male names are preferred in 14.8\%. Gray regions indicate disparities which are not significantly different from zero (18.5\% of tests).}
\label{BF-BM}
\end{figure}

\begin{figure}[t]
\includesvg[width=0.465\textwidth]{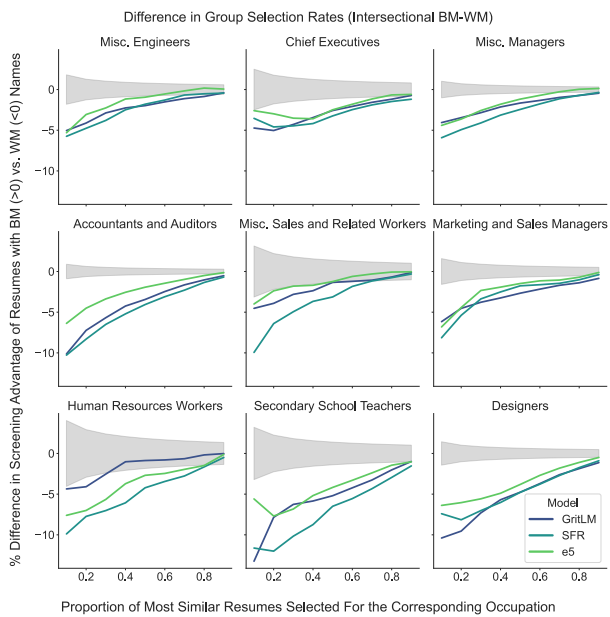}
\caption{Resumes with White male names are preferred in 100\% of tests; those with Black male names are preferred in 0\%. Gray regions indicate disparities which are not significantly different from zero (0\% of tests).}
\label{BM-WM}
\end{figure}

Larger differences appear in comparisons including Black names. Tests of resumes with White female names versus Black female names show a statistically significant preference for the former in 48.1\% of cases and the latter in only 25.9\% of cases (p$<$0.05), as shown in Figure \ref{BF-WF}. Conversely, tests of resumes with Black female names versus Black male names reveal a significant preference for the former in 66.7\% of tests and the latter in only 14.8\% of tests (p$<$0.05), as shown in Figure \ref{BF-BM}. GritLM seems to exhibit the most consistent biases against resumes with Black female names, as this is the only model which always prefers resumes with White female names and has the highest rate of preference for Black male names as well. Finally, comparisons of resumes with Black male names to those with White male names reveal the largest biases, as in Figure \ref{BM-WM}. White male names are significantly preferred in all 27 tests (p$<$0.05).

\subsection{Shorter Resumes Result in More Bias}

Overall, we find that resumes with names and title only (\textit{title-only)} lead to more biased outcomes than those with names, titles, and content (\textit{full-length}). For title-only race tests, significant differences are observed in 96.2\% of 27 bias tests, compared to 93.7\% of bias tests of full-length resumes (p$<$0.05). Of these, resumes with White names were significantly preferred in 62.9\% of tests; Black names were preferred in 33.3\%.

For title-only gender tests, significant differences are identified in 85.2\% of cases, compared to 63\% for full-length cases; this increase is entirely attributable to an increase of significant preferences for resumes with female names. Additionally, for both gender and race tests, the difference between group selection rates increases for title-only resumes. Detailed results for title-only resumes can be seen in the Appendix.

\subsection{Frequency Effects Bias Measurements}
Finally, the frequency of names also had a significant impact on outcomes. When names which had approximately equivalent frequencies in the DOLMA corpus were used, we find that resumes with Black names are preferred to those with White names in 51.9\% of 27 tests, while those with White names are preferred to those with Black names in only 22.2\% of tests. This is the reverse pattern than what is observed when names are selected based on population proportional frequencies, exemplifying model sensitivity to low level features such as name frequency. Results figures for these alternate resumes can be seen in the Appendix.

\section{Discussion}
The results of these experiments illuminate the potential for biased outcomes when using LLMs as resume screeners, as each of the three MTEs had outcomes which favor certain social groups over others. When analyzing race and gender independently, we find that the MTEs show an overall preference for resumes with White and male names, rather than preferences that align with societal patterns. For example, resumes with Black names were preferred only for the occupations Miscellaneous Managers and Secondary School Teachers; resumes with female names were preferred for Miscellaneous Engineers, Miscellaneous Sales and Related Workers, and Marketing and Sales Managers. In neither of these cases are the occupations those that have the largest proportions of workers belonging to the preferred group according to the BLS, and further investigation is needed to explain these preferences. 

The lack of correlation between population statistics and model preferences suggests that group disparities in resume screening are a consequence of default model preferences rather than occupational patterns learned during training. This contrasts to work which finds evidence of occupational biases in LLMs \citep{kotek}, and instead aligns with research suggesting that LLMs have an overall preference for certain social groups \citep{caliskan2022,  wolfe2022, ghosh-caliskan-2023-person, cheng-etal-2023-marked}. Specifically, masculine and White concepts seem to be treated as the ``default" value by models with other identities diverging from this, rather than a set of equally distinct alternatives. In our experiments, this cay be due to the fact that White and male defaults exist across many contexts, so occupation-based nuances may be weaker or lost all together. 

Intersectional results, on the other hand, do correspond more strongly to real-world discrimination in resume screening. In this setting, Black males are consistently observed to be disadvantaged relative to all other Black and White workers \citep{pager2003mark}. Our results support this; resumes with Black male names are only preferred to Black female names and White male names in 14.8\% and 0\% of bias tests, respectively.

Furthermore, the results correspond to three additional hypotheses from \citet{ghavami2013intersectional} regarding intersectionality. First, we observe that intersectional biases are not explained by the sum of race and gender biases alone. Second, the independent race bias tests align more closely with intersectional results between groups of males than groups of females. Finally, gender biases are more similar to biases between White males and females than Black males and females.

In addition to names affecting the resume screening outcomes, low-level features such as document length and name frequency also significantly altered outcomes. These manipulations both increased the amount of biased outcomes as well as which social groups were preferred, respectively. In a real-world resume screening scenario, it would be difficult to control these naturally varying features, which have an impact on LLM outcomes and may further disadvantage certain groups \citep{jones2022capturing, anil2022exploring, wolfe2021low}.  

While there are a number of factors contributing to biased outcomes in resume screening via LLMs, one naive approach to mitigation might be removing names from resumes altogether. However, resumes from real-world job seekers differ on many additional dimensions which can signal social group membership, including educational institutions, locations, and even lexical content choices. For example, \citet{parasurama2022gendered} find that resumes written by women were more likely to use words like \textit{cared} or \textit{volunteered}, while men used words like \textit{repaired} or \textit{competed}, and these differences correlated with differences in hiring outcomes. While this study manipulated names alone due to their strong associations with certain protected characteristics, LLMs used for resume screening are likely sensitive to such signals stemming from additional sources as well. 

Other approaches to bias mitigation have identified methods which can minimize nuanced social group signals from words other than names \citep{deshpande2020, parasurama2022gendered}. Additionally, other methods such as debiasing embeddings or reranking documents have also been proposed to counteract biases in hiring and retrieval settings \citep{gerritse2020effect, sundararaman2022debiasing, parasurama2021degendering}. However, these methods all rely on views of race and gender in which the ideal relationship between groups is one of sameness rather than difference. According to \citet{drage2022does}, addressing and solving inequities in resume screening mediated by AI or LLMs requires accounting for structural power imbalances that underpin the conceptualization and use of these tools. 

The primary limitations of this study were data-related. The resumes may diverge from those being used by real-world applicants due to the need to truncate them for computational feasibility. Furthermore, we relied on external tools to determine occupation categories for each document. The codes may not be as accurate as manual coding, which limits the conclusions which can be drawn for particular occupations.

The results presented here exemplify the risks associated with using LLMs for resume screening. They consistently replicate existing societal patterns of discrimination, further disadvantaging the groups already experiencing inequity in resume screening \citep{bertrand2004emily, pager2003mark}. Additionally, screening outcomes are highly variable based on low-level features such as name frequency and resume length. Rather than LLMs having the potential to counteract people's unconscious biases, we find these are reinforced in ways that are unpredictable and difficult to control in a real-world resume screening setting. Therefore, policy and mechanisms to comprehensively audit resume screening systems, whether proprietary or open source, are essential in order to evaluate their fairness and improve or remove these systems accordingly.

\section{Future Work}
Given the novelty of using LLMs as resume screeners, there is still much work to do in documenting their potential risks and improving their transparency in ways that can potentially help to identify and reduce bias and discrimination in hiring scenarios. Areas for future research include assessing additional MTEs, increasing the diversity of social group signals in resumes, as well as the range of social groups investigated. This study was limited to analyzing only two of the most commonly studied race (White and Black) and gender (male and female) groups via associated names. Hiring discrimination is not limited to these groups or signals, thus it will be important to investigate additional groups in order to fully quantify the risks in using LLMs for hiring. Finally, investigating realistic variations in resume length is an important direction as well.

\section{Conclusion}
We proposed using retrieval to simulate resume screening via LLMs to investigate the potential for biased outcomes. We found that the models do not exhibit occupational biases, but rather reinforce societal "defaults," such as the preference for White and male identities. Further intersectional analyses showed that Black males are at the greatest disadvantage, and three hypotheses of intersectionality were also confirmed in this setting. Finally, we investigated how features like resume length and name frequency can also impact biased outcomes.

\section*{Ethical Considerations Statement}
Resume screening using LLMs can be potentially difficult to research and audit ethically. Of primary importance is the preservation of privacy and confidentiality when using documents, such as resumes, which contain large amounts of identifiable information. Researchers interested in transparency and reproducibility must negotiate tensions between the distribution and use of documents which accurately reflect signals of identity such as race and gender as they would be present in real-world resumes, while also preserving privacy of the candidates represented by the documents. 

Additionally, audit studies primarily represent race and gender through names, which is a reductive and incomplete way of representing these facets of identity. Despite its limitations, it is currently one of the main approaches to study the impact of social groups and their protected attributes. As more sophisticated ways are developed to more accurately represent protected groups (including, but not limited to the race and gender groups investigated here), this study should be extended to investigate the impact of LLMs use as resume screeners on additional populations.

\section*{Acknowledgements}
We are grateful to the anonymous reviewers for their helpful feedback. This work was supported by the U.S. National Institute of Standards and Technology (NIST) Grant 60NANB23D194. Any opinions, findings, and conclusions or recommendations expressed in this material are those of the authors and do not necessarily reflect those of NIST.

\bibliography{aaai24.bib}

\clearpage
\newpage

\appendix
\onecolumn

\section{Appendix: Task Instructions}
\begin{table}[!h]
\small
\centering
\begin{tabular}{|l|l|}
\hline
\multicolumn{1}{|c|}{\textbf{ID}} & \multicolumn{1}{c|}{\textbf{Instruction}}                                                                                    \\ \hline
1                                 & \begin{tabular}[c]{@{}l@{}}Given a job description, retrieve resumes  that satisfy the requirements\end{tabular}           \\ \hline
2                                 & \begin{tabular}[c]{@{}l@{}}Given a job posting, retrieve resumes that meet the specifications\end{tabular}                \\ \hline
3                                 & \begin{tabular}[c]{@{}l@{}}Given a job profile, find resumes that fulfill the criteria\end{tabular}                       \\ \hline
4                                 & \begin{tabular}[c]{@{}l@{}}Given a job posting, find work histories that satisfy the requirements\end{tabular}            \\ \hline
5                                 & \begin{tabular}[c]{@{}l@{}}Given a job description, retrieve employment records that meet the specifications\end{tabular} \\ \hline
6                                 & \begin{tabular}[c]{@{}l@{}}Given a job profile, retrieve work  histories that satisfy the requirements\end{tabular}        \\ \hline
7                                 & \begin{tabular}[c]{@{}l@{}}Given a job profile, retrieve employment records that fulfill the criteria\end{tabular}        \\ \hline
8                                 & \begin{tabular}[c]{@{}l@{}}Given a job posting, retrieve resumes that satisfy the requirements\end{tabular}               \\ \hline
9                                 & \begin{tabular}[c]{@{}l@{}}Given a job posting, retrieve employment records that meet the specifications\end{tabular}     \\ \hline
10                                & \begin{tabular}[c]{@{}l@{}}Given a job description, retrieve work histories that fulfill the criteria\end{tabular}        \\ \hline
\end{tabular} 
\caption{Instructions appended to each job description before embedding generation.}
\label{instructions}

\end{table}

\clearpage
\newpage

\section{Appendix: Length and Frequency Plots}
\label{appendix:length}

\begin{figure*}[!h]
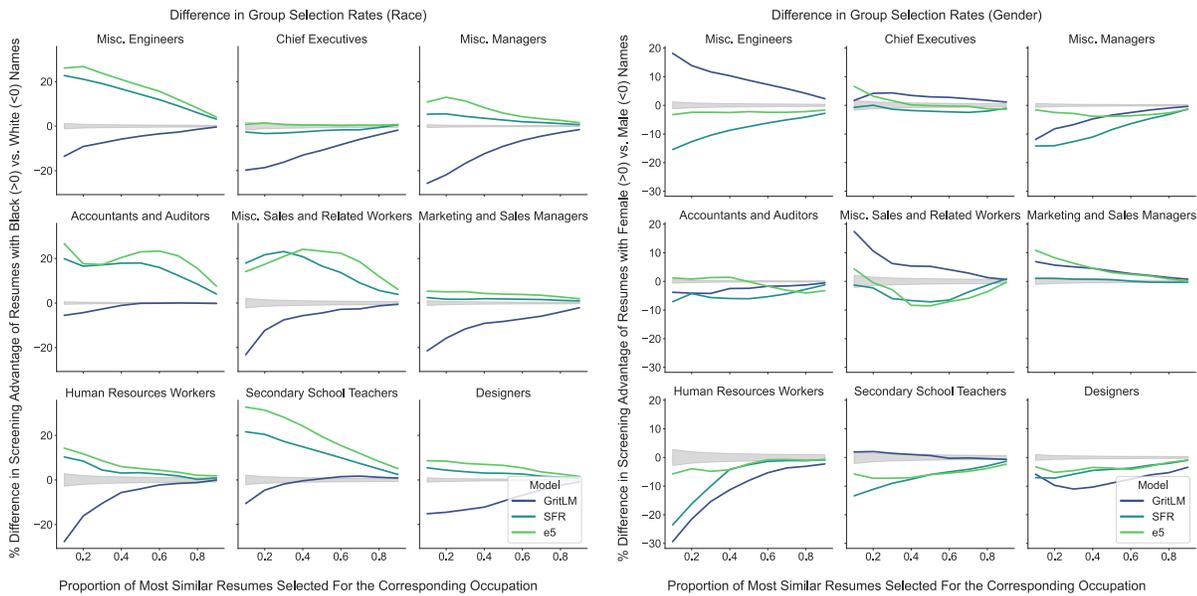

  \begin{subfigure}[t]{0.45\textwidth}
    \includesvg[width=\textwidth]{all_Rselect2_chi_jobs_len=title.svg}
    \caption{Resumes with White names were significantly preferred in 62.9\% of tests; Black names were preferred in 33.3\% of tests. No significant differences were found in 3.8\% of tests.}
    \label{race-title}
  \end{subfigure}
  \begin{subfigure}[t]{0.45\textwidth}
    \includesvg[width=\textwidth]{all_Gselect2_chi_jobs_len=title.svg}
    \caption{Resumes with male names were significantly preferred in 51.9\% of tests; female names were preferred in 33.3\% of tests. No significant differences were found in 14.8\% of tests.}
    \label{gender-title}
  \end{subfigure}\hfill
  \caption{Differences between proportion of selected groups when using resumes consisting only of names and occupation titles. Gray regions indicate ranges of group disparity which are not significantly different from zero.} 
  \label{title}
\end{figure*}

\begin{figure}[!h]
\centering
\includesvg[width=0.45\textwidth]{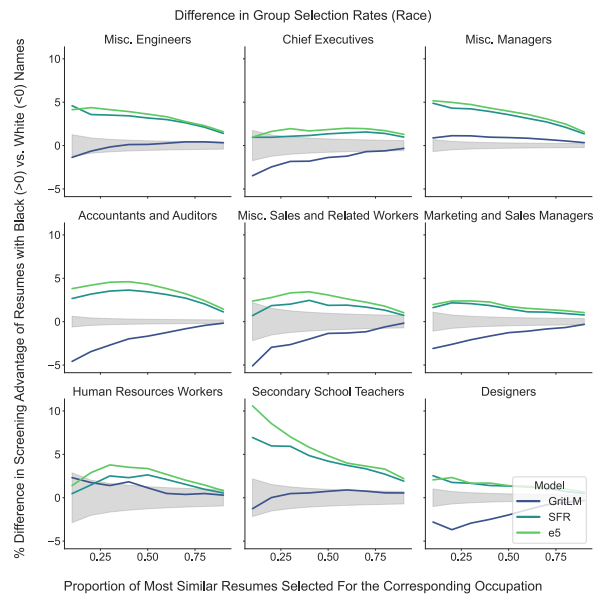}
\caption{When names chosen to augment resumes are matched using exact corpus frequencies and not based on population differences between racial groups, resumes with Black names are preferred to those with White names in 51.9\% of 27
tests, while those with White names are preferred to
those with Black names in only 22.2\% of tests. Gray regions indicate ranges of group disparity which are not significantly different from zero (25.9\% of tests).}
\label{race-unadj}
\end{figure}

\clearpage

\section{Appendix: Names} 

\begin{table*}[h!]
\centering
\begin{tabular}{|l|c|c|c|c|c|}
\hline
\textbf{\begin{tabular}[c]{@{}l@{}}First \\ Name\end{tabular}} & \textbf{\begin{tabular}[c]{@{}c@{}}DOLMA \\ First \\ Name \\ Freq.\end{tabular}} & \textbf{\begin{tabular}[c]{@{}c@{}}DOLMA \\ First\\ Name \\ Lg. Freq.\end{tabular}} & \textbf{\begin{tabular}[c]{@{}c@{}}DOLMA \\ Full\\ Name \\ Freq.\end{tabular}} & \textbf{\begin{tabular}[c]{@{}c@{}}Distinct\\ ive\\ ness\end{tabular}} & \textbf{Group} \\ \hline
Kenya                                                          & 21286328                                                                         & 16.87                                                                               & 588                                                                            & 1.80                                                                   & BF             \\
Ebony                                                          & 3921350                                                                          & 15.18                                                                               & 2109                                                                           & 1.29                                                                   & BF             \\
Chandra                                                        & 2548704                                                                          & 14.75                                                                               & 305                                                                            & 0.88                                                                   & BF             \\
Monique                                                        & 1383518                                                                          & 14.14                                                                               & 2550                                                                           & 0.87                                                                   & BF             \\
Lawanda                                                        & 1244882                                                                          & 14.03                                                                               & 121                                                                            & 1.17                                                                   & BF             \\
Asha                                                           & 1122512                                                                          & 13.93                                                                               & 140                                                                            & 0.72                                                                   & BF             \\
Aisha                                                          & 1075375                                                                          & 13.89                                                                               & 351                                                                            & 1.28                                                                   & BF             \\
Tasha                                                          & 900744                                                                           & 13.71                                                                               & 1111                                                                           & 1.10                                                                   & BF             \\
Desiree                                                        & 611231                                                                           & 13.32                                                                               & 1166                                                                           & 0.82                                                                   & BF             \\
Sheena                                                         & 549052                                                                           & 13.22                                                                               & 503                                                                            & 0.74                                                                   & BF             \\
Nisha                                                          & 444629                                                                           & 13.00                                                                               & 55                                                                             & 0.67                                                                   & BF             \\
Keisha                                                         & 411980                                                                           & 12.93                                                                               & 1029                                                                           & 1.30                                                                   & BF             \\
Tamika                                                         & 170511                                                                           & 12.05                                                                               & 1749                                                                           & 0.97                                                                   & BF             \\
Tanisha                                                        & 138490                                                                           & 11.84                                                                               & 686                                                                            & 1.83                                                                   & BF             \\
Latoya                                                         & 117619                                                                           & 11.68                                                                               & 700                                                                            & 1.09                                                                   & BF             \\
Damaris                                                        & 109196                                                                           & 11.60                                                                               & 29                                                                             & 0.79                                                                   & BF             \\
Demetria                                                       & 67394                                                                            & 11.12                                                                               & 297                                                                            & 0.90                                                                   & BF             \\
Latasha                                                        & 67215                                                                            & 11.12                                                                               & 321                                                                            & 1.71                                                                   & BF             \\
Latrice                                                        & 52295                                                                            & 10.86                                                                               & 256                                                                            & 0.71                                                                   & BF             \\
Latisha                                                        & 47547                                                                            & 10.77                                                                               & 221                                                                            & 1.57                                                                   & BF             \\ \hline
Jackson                                                        & 48887180                                                                         & 17.71                                                                               & 9090                                                                           & 0.67                                                                   & BM             \\
Abdul                                                          & 10664315                                                                         & 16.18                                                                               & 933                                                                            & 0.82                                                                   & BM             \\
Ahmad                                                          & 5616918                                                                          & 15.54                                                                               & 223                                                                            & 0.98                                                                   & BM             \\
Mohammad                                                       & 4590241                                                                          & 15.34                                                                               & 3                                                                              & 1.34                                                                   & BM             \\
Jerome                                                         & 4425676                                                                          & 15.30                                                                               & 17153                                                                          & 0.82                                                                   & BM             \\
Dante                                                          & 4035196                                                                          & 15.21                                                                               & 1253                                                                           & 0.72                                                                   & BM             \\
Lamar                                                          & 4016351                                                                          & 15.21                                                                               & 8959                                                                           & 1.12                                                                   & BM             \\
Jamal                                                          & 2251073                                                                          & 14.63                                                                               & 8092                                                                           & 1.27                                                                   & BM             \\
Desmond                                                        & 2241511                                                                          & 14.62                                                                               & 2914                                                                           & 0.90                                                                   & BM             \\
Darius                                                         & 1933306                                                                          & 14.47                                                                               & 3392                                                                           & 1.13                                                                   & BM             \\
Darrell                                                        & 1723127                                                                          & 14.36                                                                               & 8765                                                                           & 0.71                                                                   & BM             \\
Leroy                                                          & 1666967                                                                          & 14.33                                                                               & 4009                                                                           & 1.02                                                                   & BM             \\
Tyrone                                                         & 1662252                                                                          & 14.32                                                                               & 8664                                                                           & 1.72                                                                   & BM             \\
Lamont                                                         & 1367455                                                                          & 14.13                                                                               & 1926                                                                           & 0.88                                                                   & BM             \\
Cedric                                                         & 1274706                                                                          & 14.06                                                                               & 1981                                                                           & 0.99                                                                   & BM             \\
Terrell                                                        & 1204223                                                                          & 14.00                                                                               & 3379                                                                           & 1.06                                                                   & BM             \\
Jermaine                                                       & 803479                                                                           & 13.60                                                                               & 1969                                                                           & 1.42                                                                   & BM             \\
Darnell                                                        & 650613                                                                           & 13.39                                                                               & 4256                                                                           & 0.93                                                                   & BM             \\
Demetrius                                                      & 510185                                                                           & 13.14                                                                               & 2932                                                                           & 0.89                                                                   & BM             \\
Dewayne                                                        & 164800                                                                           & 12.01                                                                               & 1184                                                                           & 0.97                                                                   & BM   \\ \hline         
\end{tabular}
\caption{Complete list of Black first names used to augment resumes.}
\label{Bnames}
\end{table*}

\begin{table*}[h!]
\centering
\begin{tabular}{|l|c|c|c|c|c|}
\hline
\textbf{\begin{tabular}[c]{@{}l@{}}First \\ Name\end{tabular}} & \textbf{\begin{tabular}[c]{@{}c@{}}DOLMA \\ First \\ Name \\ Freq.\end{tabular}} & \textbf{\begin{tabular}[c]{@{}c@{}}DOLMA \\ First\\ Name \\ Lg. Freq.\end{tabular}} & \textbf{\begin{tabular}[c]{@{}c@{}}DOLMA \\ Full\\ Name \\ Freq.\end{tabular}} & \textbf{\begin{tabular}[c]{@{}c@{}}Distinct\\ ive\\ ness\end{tabular}} & \textbf{Group} \\
\hline
May & 280705894 & 19.45 & 4494 & 0.66 & WF \\
Hope & 57316784 & 17.86 & 3952 & 1.07 & WF \\
Sarah & 30459680 & 17.23 & 32882 & 1.06 & WF \\
Morgan & 24189785 & 17.00 & 12509 & 0.76 & WF \\
Jane & 23908445 & 16.99 & 23842 & 1.48 & WF \\
Beth & 23245633 & 16.96 & 9648 & 1.29 & WF \\
Anna & 21430782 & 16.88 & 14431 & 1.03 & WF \\
Liz & 12614267 & 16.35 & 9886 & 0.92 & WF \\
Ruth & 12278798 & 16.32 & 16539 & 1.01 & WF \\
Lucy & 10714752 & 16.19 & 9641 & 0.79 & WF \\
Amber & 9479469 & 16.06 & 4411 & 1.09 & WF \\
Ellen & 9289799 & 16.04 & 5887 & 1.46 & WF \\
Lily & 9184754 & 16.03 & 3309 & 0.82 & WF \\
Janet & 7588641 & 15.84 & 6386 & 0.85 & WF \\
Clara & 6967167 & 15.76 & 2840 & 0.82 & WF \\
Erin & 6670765 & 15.71 & 5198 & 1.53 & WF \\
Courtney & 4494088 & 15.32 & 14035 & 1.16 & WF \\
Heidi & 3631820 & 15.11 & 4017 & 1.41 & WF \\
Stacy & 2810977 & 14.85 & 2844 & 1.37 & WF \\
Kristine & 917872 & 13.73 & 783 & 1.30 & WF \\ \hline
John & 221590163 & 19.22 & 643330 & 0.89 & WM \\
Joe & 59609012 & 17.90 & 78153 & 1.14 & WM \\
Kevin & 31483073 & 17.26 & 45484 & 0.77 & WM \\
Fred & 25530058 & 17.06 & 23114 & 0.94 & WM \\
Grant & 24524525 & 17.02 & 56117 & 1.35 & WM \\
Luke & 21934606 & 16.90 & 29502 & 1.32 & WM \\
Howard & 21647437 & 16.89 & 9268 & 0.95 & WM \\
Danny & 12508368 & 16.34 & 61760 & 0.90 & WM \\
Pete & 12512264 & 16.34 & 25971 & 1.49 & WM \\
Nicholas & 10735251 & 16.19 & 7728 & 0.84 & WM \\
Brent & 9533529 & 16.07 & 6825 & 1.49 & WM \\
Stuart & 9307204 & 16.05 & 12145 & 1.18 & WM \\
Arnold & 9134145 & 16.03 & 2098 & 1.29 & WM \\
Milton & 7828179 & 15.87 & 6292 & 0.71 & WM \\
Harold & 7003824 & 15.76 & 8235 & 0.91 & WM \\
Wesley & 6510149 & 15.69 & 5227 & 0.88 & WM \\
Corey & 4476722 & 15.31 & 12714 & 0.69 & WM \\
Theodore & 3594280 & 15.09 & 1777 & 0.72 & WM \\
Stevie & 2907782 & 14.89 & 7004 & 0.66 & WM \\
Huey & 920303 & 13.73 & 332 & 0.69 & WM \\ \hline
\end{tabular}
\caption{Complete list of White first names used to augment resumes. Their corpus frequencies are 5.5 times more frequent than corresponding Black names, reflecting US population differences between numbers of White and Black people.}
\label{Wnames}
\end{table*}

\begin{table*}[h!]
\centering
\begin{tabular}{|l|c|c|c|c|c|}
\hline
\textbf{\begin{tabular}[c]{@{}l@{}}First \\ Name\end{tabular}} & \textbf{\begin{tabular}[c]{@{}c@{}}DOLMA \\ First \\ Name \\ Freq.\end{tabular}} & \textbf{\begin{tabular}[c]{@{}c@{}}DOLMA \\ First\\ Name \\ Lg. Freq.\end{tabular}} & \textbf{\begin{tabular}[c]{@{}c@{}}DOLMA \\ Full\\ Name \\ Freq.\end{tabular}} & \textbf{\begin{tabular}[c]{@{}c@{}}Distinct\\ ive\\ ness\end{tabular}} & \textbf{Group} \\ \hline
Virginia & 55841100 & 17.84 & 11137 & 1.05 & =WF \\
Katie & 10648070 & 16.18 & 9254 & 1.66 & =WF \\
Daisy & 5607637 & 15.54 & 1295 & 1.06 & =WF \\
Danielle & 4578506 & 15.34 & 9107 & 0.72 & =WF \\
Courtney & 4494088 & 15.31 & 14035 & 1.16 & =WF \\
Laurie & 4040648 & 15.21 & 5787 & 1.01 & =WF \\
Brittany & 3991685 & 15.20 & 6629 & 1.46 & =WF \\
Christy & 2250120 & 14.63 & 2440 & 1.22 & =WF \\
Melinda & 2253175 & 14.63 & 1397 & 1.13 & =WF \\
Terri & 1942964 & 14.48 & 3429 & 0.82 & =WF \\
Lillian & 1720315 & 14.36 & 1724 & 1.00 & =WF \\
Ethel & 1695992 & 14.34 & 1899 & 1.14 & =WF \\
Elmer & 1626446 & 14.30 & 791 & 0.74 & =WF \\
Mildred & 1346379 & 14.11 & 4324 & 1.34 & =WF \\
Marlene & 1248880 & 14.04 & 715 & 0.66 & =WF \\
Aimee & 1185761 & 13.99 & 564 & 0.76 & =WF \\
Sherri & 808398 & 13.60 & 1755 & 1.00 & =WF \\
Vickie & 631365 & 13.36 & 1241 & 1.20 & =WF \\
Aileen & 499873 & 13.12 & 148 & 0.87 & =WF \\
Rebeca & 158389 & 11.97 & 24 & 1.30 & =WF \\ \hline
Daniel & 49113143 & 17.71 & 27976 & 1.06 & =WM \\
Spencer & 10599899 & 16.18 & 11108 & 0.90 & =WM \\
Shawn & 5632418 & 15.54 & 20698 & 0.80 & =WM \\
Teddy & 4616032 & 15.35 & 3935 & 0.76 & =WM \\
Ricky & 4401901 & 15.30 & 58492 & 0.89 & =WM \\
Randall & 3993460 & 15.20 & 11979 & 0.87 & =WM \\
Cary & 3872735 & 15.17 & 13868 & 1.17 & =WM \\
Ernie & 2216293 & 14.61 & 1852 & 0.75 & =WM \\
Denny & 2287179 & 14.64 & 634 & 1.13 & =WM \\
Sylvester & 1950442 & 14.48 & 9766 & 0.82 & =WM \\
Mathew & 1657498 & 14.32 & 1903 & 1.24 & =WM \\
Scotty & 1508539 & 14.23 & 511 & 1.64 & =WM \\
Donnie & 1375895 & 14.13 & 2812 & 0.84 & =WM \\
Timmy & 1247804 & 14.03 & 1155 & 1.58 & =WM \\
Donny & 1162977 & 13.97 & 1174 & 1.02 & =WM \\
Johnnie & 985966 & 13.80 & 1676 & 0.96 & =WM \\
Earnest & 829440 & 13.63 & 543 & 0.67 & =WM \\
Johnathan & 572841 & 13.26 & 7250 & 1.01 & =WM \\
Bennie & 421825 & 12.95 & 564 & 0.70 & =WM \\
Wilbert & 221959 & 12.31 & 312 & 0.87 & =WM \\ \hline
\end{tabular}
\caption{Complete list of White first names used to augment resumes for supplemental analysis. Their corpus frequencies are roughly equal to corresponding Black names.}
\label{=Wnames}
\end{table*}
\clearpage

\label{appendix:b}

\section{Appendix: Document Lengths} \label{lengths-appendix}

\begin{figure}[h]
\centering
\includesvg[width=0.465\textwidth]{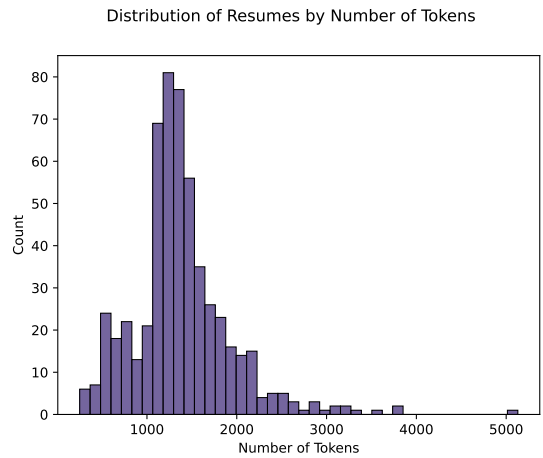}
\caption{Before resumes were truncated, they have a wide range of lengths. Average length is 1,154 tokens.}
\label{resume-len}
\end{figure}

\begin{figure}[h]
\centering
\includesvg[width=0.465\textwidth]{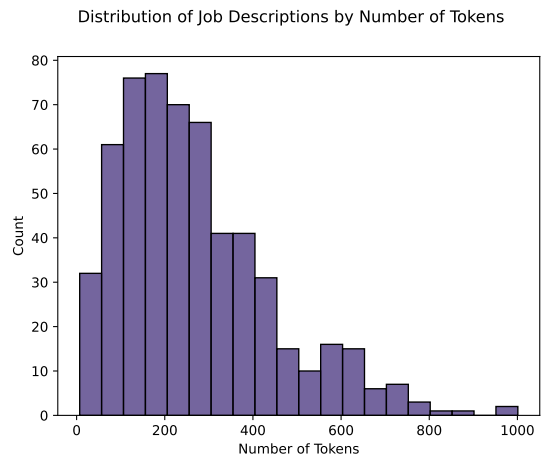}
\caption{Job descriptions are shorter than resumes, and they are not truncated as part of pre-processing.}
\label{description-len}
\end{figure}
\clearpage

\section{Appendix: Detailed Instructions Results} \label{task-appendix}

\begin{figure*}[h]
\includesvg[width=\textwidth]{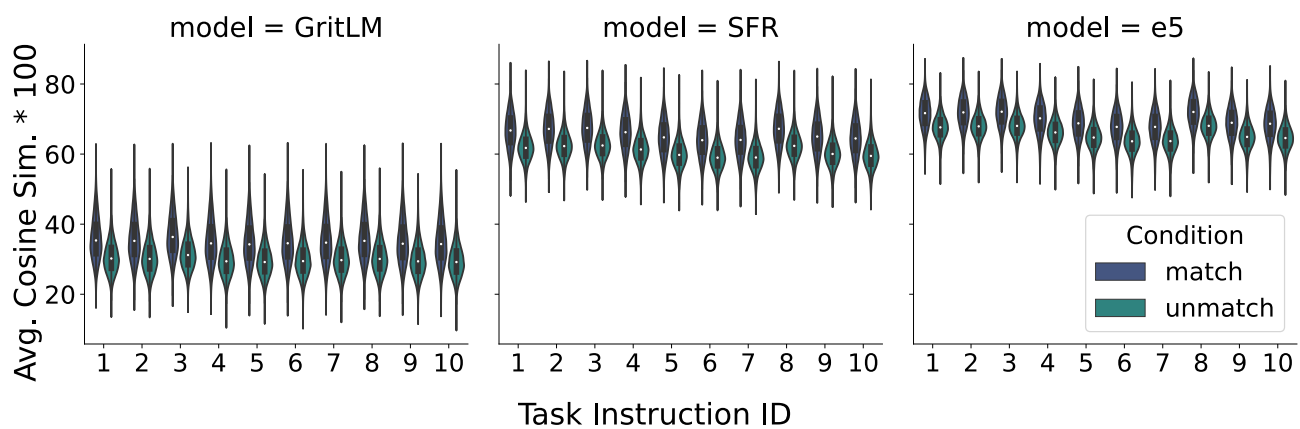}
\caption{For every model and task instruction corresponding to those in Table \ref{instructions}, resumes which belong to the same occupation category as a given job description (\textit{match}) have significantly higher cosine similarities than those which belong to different occupation categories (\textit{unmatch}) (p$<$0.0001).}
\label{task-scores}
\end{figure*}
\clearpage

\section*{Appendix: Detailed Cosine Similarity Results} \label{cosine-appendix}
\begin{figure}[h]
\centering
\includesvg[width=0.465\textwidth]{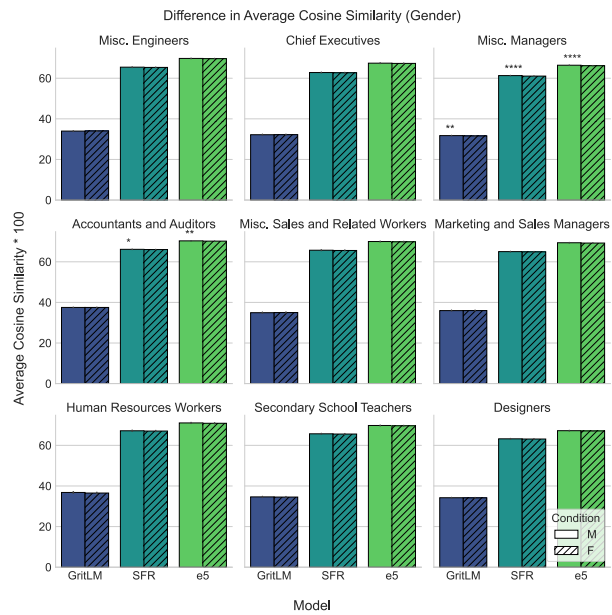}
\caption{Cosine similarities are significantly larger for male Accountants and Auditors and Misc. Managers. Significant differences are indicated with asterisks: (*p$<$0.05, **p$<$0.01, ***p$<$0.001, ****p$<$0.0001).}
\label{gender-cosine-full}
\end{figure}

\begin{figure}[h]
\centering
\includesvg[width=0.465\textwidth]{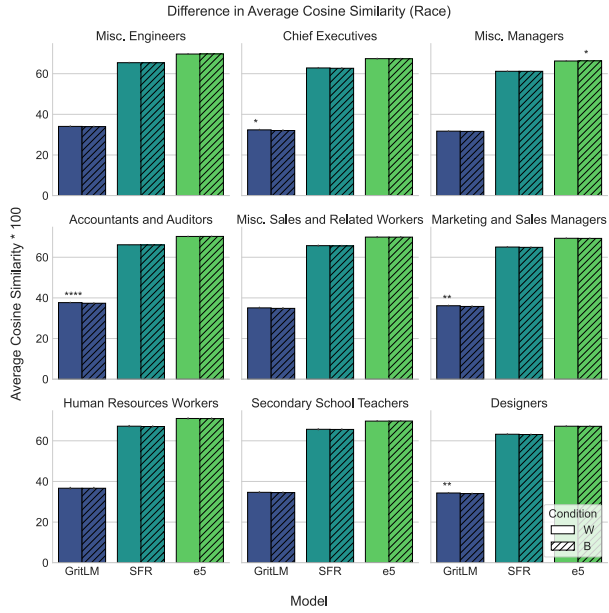}
\caption{Cosine similarities are significantly higher for White Chief Executives, Marketing and Sales Managers, and Designers; resumes which Black names have higher cosine similarity for Misc. Managers. Significant differences are indicated with asterisks: (*p$<$0.05, **p$<$0.01, ***p$<$0.001, ****p$<$0.0001).}
\label{race-cosine-full}
\end{figure}

\begin{figure}
\centering
\includesvg[width=0.465\textwidth]{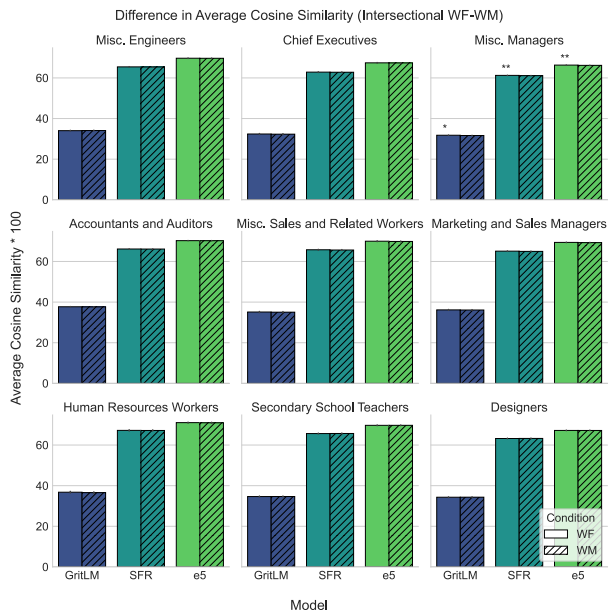}
\caption{Cosine similarities for resumes with White female names are significantly higher than White male names for Misc. Managers. Significant differences are indicated with asterisks: (*p$<$0.05, **p$<$0.01, ***p$<$0.001, ****p$<$0.0001).}
\label{WF-WM-cosine-full}
\end{figure}

\begin{figure}
\centering
\includesvg[width=0.465\textwidth]{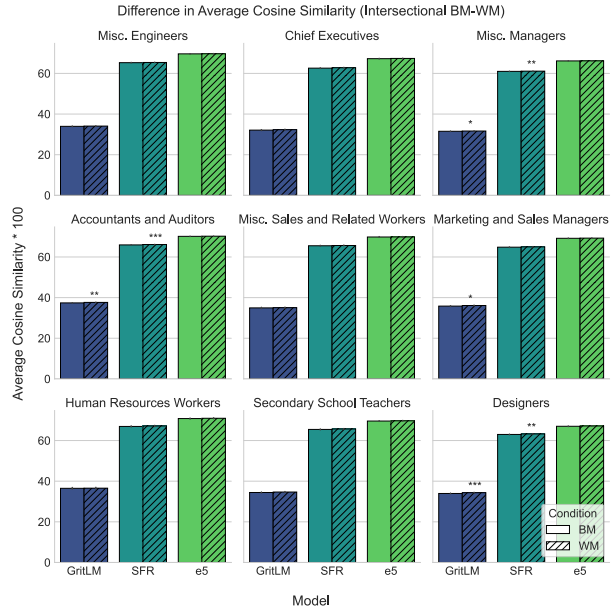}
\caption{Cosine similarities for resumes with White male names are significantly higher than Black male names for Misc. Managers, Marketing and Sales Managers, Accountants and Auditors, and Designers. Significant differences are indicated with asterisks: (*p$<$0.05, **p$<$0.01, ***p$<$0.001, ****p$<$0.0001).}
\label{BM-WM-cosine-full}
\end{figure}

\begin{figure}
\centering
\includesvg[width=0.465\textwidth]{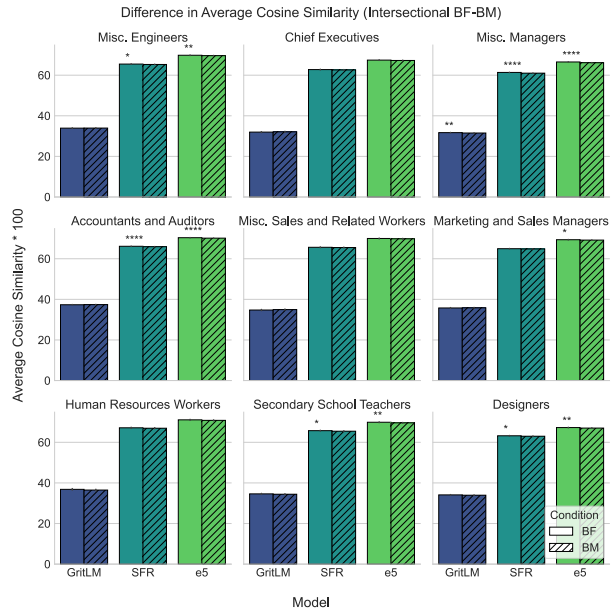}
\caption{Cosine similarities for resumes with Black female names are significantly higher than Black male names for Misc. Managers, Misc. Engineers, Marketing and Sales Managers, Accountants and Auditors, Secondary School Teachers, and Designers. Significant differences are indicated with asterisks: (*p$<$0.05, **p$<$0.01, ***p$<$0.001, ****p$<$0.0001).}
\label{BF-BM-cosine-full}
\end{figure}

\begin{figure}
\centering
\includesvg[width=0.465\textwidth]{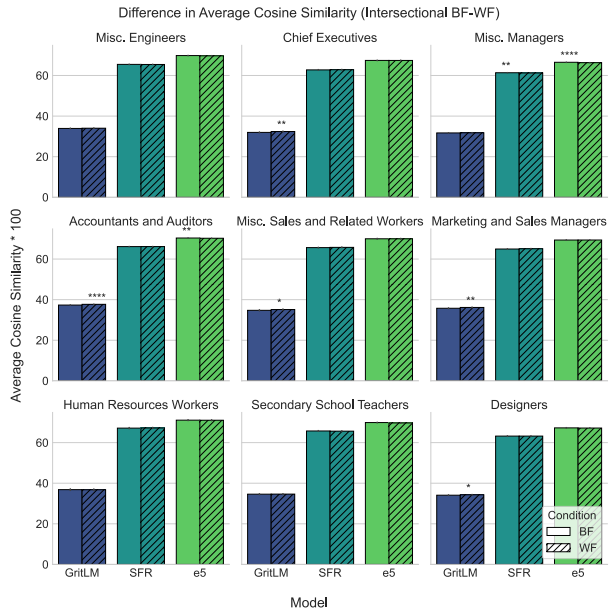}
\caption{Cosine similarities for resumes with Black female names are significantly higher than White female names for Misc. Managers. Resumes with White female names have significantly higher cosine similarities for Marketing and Sales Managers, Designers, Misc. Sales and Related Workers, and Chief Executives. Both sets of names have higher cosine similarities for Accountants and Auditors depending on the model. Significant differences are indicated with asterisks: (*p$<$0.05, **p$<$0.01, ***p$<$0.001, ****p$<$0.0001).}
\label{WF-BF-cosine-full}
\end{figure}

\begin{figure}
\centering
\includesvg[width=0.465\textwidth]{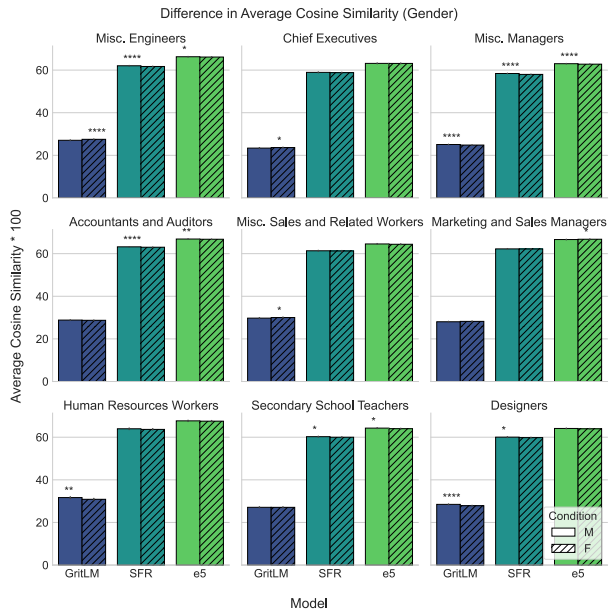}
\caption{When resumes are title-only, those with male names have higher cosine similarities for Designers, Secondary School Teachers, Misc. Managers, Accountants and Auditors, and Human Resources Workers. Those with female names have higher cosine similarity for Chief Executives and Misc. Sales and Related Workers. Both sets of names have higher cosine similarities for Misc. Engineers depending on the model. Significant differences are indicated with asterisks: (*p$<$0.05, **p$<$0.01, ***p$<$0.001, ****p$<$0.0001).}
\label{G-cosine-title}
\end{figure}

\begin{figure}
\centering
\includesvg[width=0.465\textwidth]{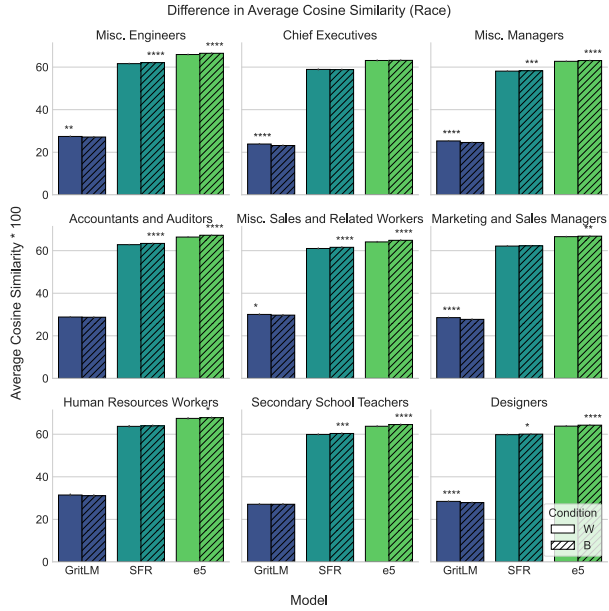}
\caption{When resumes are title-only, those with White names have higher cosine similarities for Chief Executives. Those with Black names have higher cosine similarity for Secondary School Teachers, Accountants and Auditors, Misc. Sales and Related Workers, and Human Resources Workers. Both sets of names have higher cosine similarities for Misc. Engineers, Misc. Managers, Marketing and Sales Managers, and Designers depending on the model. Significant differences are indicated with asterisks: (*p$<$0.05, **p$<$0.01, ***p$<$0.001, ****p$<$0.0001).}
\label{R-cosine-title}
\end{figure}

\begin{figure}
\centering
\includesvg[width=0.465\textwidth]{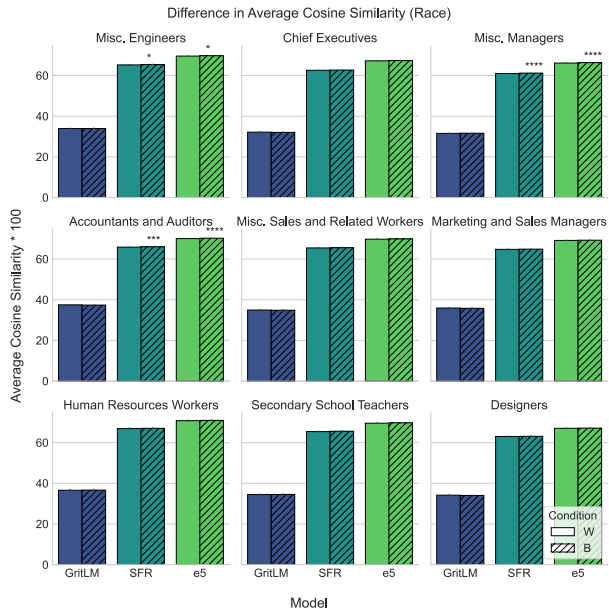}
\caption{When resumes have names with approximately equal corpus frequencies, those with Black names have higher cosine similarity for Misc. Engineers, Misc. Managers, and Accountants and Auditors. Significant differences are indicated with asterisks: (*p$<$0.05, **p$<$0.01, ***p$<$0.001, ****p$<$0.0001).}
\label{R-cosine-unadj}
\end{figure}

\end{document}